# 'Beyond quantum theory: a realist psycho-biological interpretation of reality' revisited

Brian D. Josephson

*Cavendish Laboratory, Madingley Road, Cambridge CB3 0HE, UK*

**Abstract**

It is hypothesised, following Conrad et al. (1988) that quantum physics is not the ultimate theory of nature, but merely a theoretical account of the phenomena manifested in nature under particular conditions. These phenomena parallel cognitive phenomena in biosystems in a number of ways and are assumed to arise from related mechanisms. Quantum and biological accounts are complementary in the sense of Bohr and quantum accounts may be incomplete. In particular, following ideas of Stapp, 'the observer' is a system that, while lying outside the descriptive capacities of quantum mechanics, creates observable phenomena such as wave function collapse through its probing activities. Better understanding of such processes may pave the way to new science.

Michael Conrad was an unusually gifted scientist. My experience with him was that if one had a question about anything one could go and ask him about it and get back a clear explanation of the issue concerned, no matter what field it belonged to. And if one was working on an idea of one's own and wanted some feedback, he would always come back with deep insights.

I will leave to others in this volume the task of explaining his many innovative ideas, and focus here on some specific ideas that we worked on together. One of these was the idea from Eastern Philosophy that in certain states of consciousness the *subjective* states of the mind, irrespective of learning, closely reflect *objective* reality, a state of affairs contrary to that of the usual assumption, whereby the contents of the mind reflect objective reality purely as a consequence of what one has learnt about it. Such an idea had been discussed by Fritjof Capra in his book the Tao of Physics (Capra 1983), concerned with the deep parallels that appear to exist between patterns found in objective reality as revealed by modern science, and patterns found in deeper personal experiences as revealed by meditation or mystical experience and reported by the mystics. A related theoretical idea, based on Whitehead's process philosophy, was developed by Stapp (1982, 1985). This is the idea that reality evolves by a mind-like process, decisions made by this process being apparent in the context of ordinary physics as the collapse of the wave function. In our Urbino conference paper (Conrad et al. 1988) we tried to take this idea further (see Table 1), proposing a number of logical correspondences between the two modes of description (in the original paper we called the right hand side biological, since we regarded phenomena such as signals, decisions and regulation as characteristically biological, a theme developed in more detail in Josephson and Conrad (1992)):

*[table 1 about by here]*

The details of quantum physics and biology are very different, but we argued that they might nevertheless be derivative of some common underlying subtler background process, in the same way that waves and particles emerge from a common subtler domain, that of quantum mechanics, and in some cases share certain features such as propagation along a trajectory. Quantum mechanics would then be the specific theory that emerges as a good description in some domain of nature, whilst more biological accounts would be relevant in some other phenomenal domain. We thus envisaged the possibility, highlighted in some of the writings of Bohr (1958), that biological and quantum accounts of nature might, like the wave and particle accounts, of certain phenomena, be complementary rather than, as with the conventional view, the first being entirely derivative of the latter.

We finished our paper with considerations of knowability (in which discussion our coauthor, Dipankar Home, played a major role), it being our view that the form of a scientific domain is very much influenced by its paradigm. Biology concerns itself largely with processes, while quantum mechanics is concerned fundamentally with quantifiability. As already noted, these aspects may be complementary and also incompatible. Quantum mechanics achieves its quantitative aspects by an averaging process, but this may lead to neglecting characteristics of individual cases which may be relevant in the case of a biosystem, provided we are prepared to recognise the uniqueness of the individual case instead of treating all cases of a class as if they were the same. This may point to a fundamental inadequacy in the quantum point of view, as we illustrated by consideration of a classical gas where the options exist for statistical or deterministic accounts, there being an epistemology acknowledging only statistical properties or properties described in terms of macroscopic fields, and also an epistemology involving an entirely new area of knowledge relating to individual particles. One might then see the ruling role of quantum mechanics as an artefact of our scientific culture which, in the domain where quantum indeterminism is of importance, has chosen to be blind as regards individual cases, concern with understanding the many possibilities having distracted us from concern with what the one that actually corresponds to reality for us. Josephson and Pallikari-Viras (1991) suggested that this difference may be important in connection with the understanding of paranormal phenomena.

It is of interest to take these concepts further by bringing in Stapp's more recent ideas (Stapp 2001), which involve explicitly the role of the observer. In this formulation, based on the von Neumann interpretation of quantum mechanics, the observer *poses questions of nature*, which in the process of answering them becomes better defined. In Stapp's words, a "mental event" occurs that "grasps a whole unit of structural information, and injects it into the quantum state of the universe". Quantum mechanics does not itself indicate what questions are asked.

A way of interpreting this situation, consistent with the view taken by Michael Conrad, Dipankar Home and myself, is that the entity that poses the questions simply falls outside the scope of the quantum paradigm in the way that atoms and molecules fall outside the scope of a macroscopic account of a gas. The atoms and molecules do however show their presence in phenomena such as fluctuations even when they cannot be detected directly. In the same way, the observer who asks questions of nature and creates an 'orchestrated reduction' (Hameroff and Penrose 1996) as a result may be something that lies outside the scope of descriptions of the paradigm. But as far as the observer himself, who may be asking a question (or wanting to know something) is concerned, it is not a convenient fiction but reality.

These issues may be addressable in terms of our "beyond quantum theory" ideas, if we identify observers with something of the nature of an organism or cooperating group. As Rosen (1991, 1999) has noted, in the biological realm we may have to think in terms of causes and effects rather than states and their dynamics. Stapp's ideas fit well into such a picture, the observer, who is outside quantum mechanics, being one of the causes of effects within this descriptive domain. The conclusion then is that Stapp's observers fall outside the ability of quantum mechanics to characterise, but not in any way essentially beyond our ability to understand them and describe them in alternative ways. Science needs to try to understand the observer, and to respond vigorously to "the challenge of consciousness research" (Josephson and Rubik 1992). Physics, in advocating quantum mechanics as a basis for a "theory of everything", may have moved too fast towards a too tempting conclusion, and thrown out the crucial and subtle intelligence of the observer as a part of this process.

| Language of quantum physics | <=> | Language of information processing |
|---|---|---|
| quantum subsystem, describable by a state vector | <=> | signal or form |
| particle type | <=> | type of signal or form |
| state vector representing a specific possibility | <=> | signal representing a specific possibility |
| collapse of state vector | <=> | decision process |
| measuring instrument determining state of subsystem | <=> | structures which determine and regulate signals or forms |

Table 1. Proposed identification of entities described in terms of the respective frames of reference of the quantum physicist and the biologist. `return to text`